# *Achieving transparency with plasmonic and metamaterial coatings*

*Andrea Alù[1,2], and Nader Engheta[1,*]*
*[1]Dept. of Electrical and Systems Engineering, University of Pennsylvania, Philadelphia, 19104 PA, USA*
*[2]Dept. of Applied Electronics, University of Roma Tre, Rome, Italy*

***Abstract:*** The possibility of using plasmonic and metamaterial covers to drastically reduce the total scattering cross section of spherical and cylindrical objects is discussed. While it is intuitively expected that increasing the physical size of an object may lead to an increase in its overall scattering cross section, here we see how a proper design of these *lossless* metamaterial covers near their plasma resonance may induce a dramatic drop in the scattering cross section, making these object nearly "invisible" or "transparent" to an outside observer – a phenomenon with obvious applications for low-observability and non-invasive probe design. Physical insights into this phenomenon and some numerical results are provided.

## INTRODUCTION

Achieving "invisibility" or "low observability" has been the subject of extensive studies in the physics and engineering communities for decades. The use of absorbing screens (e.g., [1]) and anti-reflection coatings (e.g., [2]) to diminish the scattering or the reflection from objects, for example, are common in several applications. If the former require absorption, and therefore power dissipation, the latter are suitable primarily for planar or nearly-planar transparent objects. In this paper, however, we discuss how low-loss (in the limit even no-loss) passive covers might be utilized in order to drastically reduce the scattering from spherical and cylindrical objects without requiring high dissipation, but relying on a completely different mechanism.

For this purpose, we consider here the use of materials with negative or low electromagnetic constitutive parameters, e.g., certain metals near their plasma frequency or metamaterials with negative parameters. Several noble metals achieve this requirement for their electrical permittivity at the infra-red (IR) or visible regimes, even with reasonably low losses [3]-[4]. At lower frequencies, moreover, artificial materials and metamaterials may be properly synthesized to meet similar requirements at the desired frequency (see references in [5]). At microwaves, for instance, the technology of artificial materials with unusual electromagnetic properties has a long tradition in the engineering community. Materials with low-positive, near-zero or negative relative permittivity have been synthesized by embedding arrays of thin wires in a host material [6]-[8], and their properties have been exploited in different microwave applications over the years, e.g., for highly directive antennas [9]-[10] or more recently for anomalous tunneling and transmission (e.g., [11]-[12]). By a similar principle, arrays of resonating loops or split-ring resonators affect in an analogous way the permeability of the bulk material and have been studied in the last decade [13]-[14]. The combination of these inclusions in composite *meta*materials with a negative index of refraction [15] has recently attracted a great deal of attention in the scientific community.

The idea of utilizing low-permittivity coatings to induce “invisibility” has been suggested in the quasi-static limit for spherical objects (i.e., for the case of a sphere being small compared with the operating wavelength) and for the higher-order terms in power series expansion [16, p. 149-150], [17]-[19]. Here, however, we explore fully the case of plasmonic and metamaterial covers with negative or low-permittivity and/or permeability materials in the dynamic *full-wave* scattering case (i.e., the sphere does not need to be electrically *very* small), providing some new physical insights and possibility of optimized designs for reducing the total scattering cross section of spherical objects whose dimensions are comparable with the wavelength of operation.

## THEORETICAL ANALYSIS

Let us consider a spherical scatterer of radius $a$, composed of a homogeneous material with permittivity $\varepsilon$ and permeability $\mu$, surrounded by free space (with constitutive parameters $\varepsilon_0$ and $\mu_0$). (As will be mentioned later, in the limit this sphere can also be a perfect electric conductor.) Would it be possible to cover such an object with a lossless (or low-loss) spherical shell (with $\varepsilon_c$, $\mu_c$ and radius $a_c > a$) in such a way that the total scattering cross section drastically diminishes, even though its overall physical size has evidently increased? Here we explore the conditions under which this may be possible.

* To whom correspondence should be addressed, engheta@ee.upenn.edu

The geometry of the problem is depicted in Fig. 1. Let the illuminating wave be represented by an $e^{-i\omega t}$ – monochromatic plane wave. All the constitutive parameters may in general be complex at the frequency of interest $\omega$, taking into account possible losses. Adopting a standard Mie expansion, one can write the scattered field from a plane wave as the well-known discrete sum of spherical harmonics with complex amplitudes $c_n^{TE}$ and $c_n^{TM}$ ($n \geq 1$ being an integer), respectively for the TE and TM spherical waves (see e.g. [20]). Their expressions for the problem at hand may be written in the convenient form [21]:

$$c_n^{TE} = -\frac{U_n^{TE}}{U_n^{TE} + iV_n^{TE}}, \quad c_n^{TM} = -\frac{U_n^{TM}}{U_n^{TM} + iV_n^{TM}}, \tag{1}$$

where the functions $U_n$ and $V_n$ are real valued in the limit of no ohmic (material) losses, and in the TM polarization are given by the determinants:

$$U_n^{TM} = \begin{vmatrix} j_n(ka) & j_n(k_c a) & y_n(k_c a) & 0 \\ [ka\, j_n(ka)]'/\varepsilon & [k_c a\, j_n(k_c a)]'/\varepsilon_c & [k_c a\, y_n(k_c a)]'/\varepsilon_c & 0 \\ 0 & j_n(k_c a_c) & y_n(k_c a_c) & j_n(k_0 a_c) \\ 0 & [k_c a_c\, j_n(k_c a_c)]'/\varepsilon_c & [k_c a_c\, y_n(k_c a_c)]'/\varepsilon_c & [k_0 a_c\, j_n(k_0 a_c)]'/\varepsilon_0 \end{vmatrix}, \tag{2}$$

$$V_n^{TM} = \begin{vmatrix} j_n(ka) & j_n(k_c a) & y_n(k_c a) & 0 \\ [ka\, j_n(ka)]'/\varepsilon & [k_c a\, j_n(k_c a)]'/\varepsilon_c & [k_c a\, y_n(k_c a)]'/\varepsilon_c & 0 \\ 0 & j_n(k_c a_c) & y_n(k_c a_c) & y_n(k_0 a_c) \\ 0 & [k_c a_c\, j_n(k_c a_c)]'/\varepsilon_c & [k_c a_c\, y_n(k_c a_c)]'/\varepsilon_c & [k_0 a_c\, y_n(k_0 a_c)]'/\varepsilon_0 \end{vmatrix}. \tag{3}$$

Analogous expressions for the TE polarization are obtained after substituting $\varepsilon$ with $\mu$ in (2) and (3). In these formulas $k \equiv \omega\sqrt{\varepsilon\,\mu}$, $k_c \equiv \omega\sqrt{\varepsilon_c\,\mu_c}$ and $k_0 \equiv \omega\sqrt{\varepsilon_0\,\mu_0}$ are the wave numbers in the three regions and $j_n(.)$, $y_n(.)$ are spherical Bessel functions.

The scattering coefficients are related to the total scattering cross section through the formula:

$$Q_s = \frac{2\pi}{|k_0|^2}\sum_{n=1}^{\infty}(2n+1)\left(\left|c_n^{TE}\right|^2 + \left|c_n^{TM}\right|^2\right). \tag{4}$$

Generally speaking, the scattering cross section $Q_s$ is determined by the scattering coefficients up to the order $n = N_{\max}$, since the successive scattering coefficients beyond $N_{\max}$ will be negligible [16]. The value of $N_{\max}$ usually increases with the physical/electrical size of the scatterer (a rule of thumb is that $N_{\max} \simeq (k_0 a_c)$ [16, p. 126]), and that is one of the reasons that larger objects generally have a wider scattering cross section. Following this observation, one may expect, intuitively, that covering an object would increase its scattering cross section, together with its physical/electrical size, since more scattering coefficients contribute in the summation. However, as will be shown below, this statement is not always true, and with judiciously choosing the cover material and its size, one may be able to make a given homogenous (or even metallic) sphere "less detectable" to an external observer.

Let us first concentrate on the case in which the scatterer is small enough to consider only the dipolar term ($n=1$) in the Mie expansion. This case is consistent with what was done before in [16]-[19]. In this case, i.e., when $k_c a_c \ll 1$, $k_0 a_c \ll 1$, $k\,a \ll 1$, the expressions in (2) and (3) are reduced to the following [21]:

$$U_n^{TM} \simeq \frac{\pi (k_0 a_c)^{n+1}}{4^n (2n+1)(n-1/2)!^2} \begin{vmatrix} 1 & 1 & -1 & 0 \\ (n+1)/\varepsilon_1 & (n+1)/\varepsilon_2 & n/\varepsilon_2 & 0 \\ 0 & \gamma^{-n} & -\gamma^{n+1} & 1 \\ 0 & (n+1)\gamma^{-n}/\varepsilon_2 & n\gamma^{n+1}/\varepsilon_2 & (n+1)/\varepsilon_0 \end{vmatrix}, \qquad (5)$$

$$V_n^{TM} \simeq (k_0 a_c)^{-n} \begin{vmatrix} 1 & 1 & -1 & 0 \\ (n+1)/\varepsilon & (n+1)/\varepsilon_c & n/\varepsilon_c & 0 \\ 0 & \gamma^{-n} & -\gamma^{n+1} & -1 \\ 0 & (n+1)\gamma^{-n}/\varepsilon_c & n\gamma^{n+1}/\varepsilon_c & n/\varepsilon_0 \end{vmatrix}, \qquad (6)$$

where $\gamma \equiv a/a_c$ is the ratio of core-shell radii. These closed-form expressions reveal interesting properties for electrically small scatterers. Firstly, as expected, for small scatterers the value of $U_n$ is small and tends to zero as $a_c \to 0$. This is consistent with the fact that usually a small scatterer has a very low scattered field. In fact, considering standard dielectrics far from their plasma frequency (all permittivities greater or equal to $\varepsilon_0$, all permeabilities equal to $\mu_0$) $c_n^{TE} \simeq 0$ and the sum in the denominator of $c_n^{TM}$ is dominated by $V_n^{TM}$, yielding the following approximate expression for $c_n^{TM}$:

$$c_n^{TM} \cong -j (k_0 a_c)^{2n+1} f_n(\gamma), \qquad (7)$$

where $f_n(\gamma)$ are positive real functions of $\gamma$. This confirms that for such tiny dielectric scatterers the scattering properties are dominated by the first-order TM term, corresponding to the radiation from an electric dipole. This also shows that covering this structure with a dielectric shell may result in increasing its scattering cross section together with its physical/electrical size (i.e., $a_c$). However, there is one possibility for which adding a cover may lead to a drastic reduction of the total scattering cross section from a small sphere, and that is when the cover parameters are chosen such that the determinant $U_n^{TM}$ in (5) may approach zero, resulting in $c_n^{TM}$, which is the dominant term in (4), to become zero. In the limit for which (5) applies, its value will be identically zero for a given order $n$, if $\gamma$ satisfies the following condition:

$$\gamma = \sqrt[2n+1]{\frac{(\varepsilon_c-\varepsilon_0)\left[(n+1)\varepsilon_c+n\varepsilon\right]}{(\varepsilon_c-\varepsilon)\left[(n+1)\varepsilon_c+n\varepsilon_0\right]}} \equiv T_{TM} . \quad (8)$$

where $T_{TM}$ is a shorthand for the $(2n+1)$-st root. Clearly, in order for $\gamma$ to have physical meaning, its value should satisfy $0 \le \gamma \le 1$. This result is consistent with the findings in the quasi-static case in [16]-[19]. In Fig. 2 the values of permittivities for which (8) can be fulfilled are reported in a contour plot, with lighter regions indicating higher values of $T_{TM}$. When $\gamma = T_{TM}$ for a given $n$, the corresponding scattering coefficient $c_n^{TM}$ becomes identically zero (in the small-sphere approximation). Similar results may be obtained for the TE polarization by replacing the permittivities with the corresponding permeabilities. One may notice that if both materials are standard materials far from their plasma frequencies (i.e., with constitutive parameters higher than those of free space) it is not possible to achieve such a condition for any value of $\gamma$. If however we are allowed to use covers with permittivity or permeability lower than the one in free space, or with negative values, which might represent metals close to their plasma frequency or alternatively low-permittivity, low-permeability, or negative-parameter metamaterials, then following Fig. 2 a proper choice of the ratio of core-shell allows bringing to zero one particular $c_n$.

If the core sphere (the one to be hidden) is dielectric (and electrically small, in order to apply (8)), then the electric dipole term $c_1^{TM}$ is the dominant term in scattering, and all the other terms in (4) are negligible. In this case, from (8) the condition for the cover radius becomes:

$$a = \sqrt[3]{\frac{(\varepsilon_c-\varepsilon_0)\left[2\varepsilon_c+\varepsilon\right]}{(\varepsilon_c-\varepsilon)\left[2\varepsilon_c+\varepsilon_0\right]}}\, a_c , \quad (9)$$

which is consistent with the formula derived in [16, p. 150] in the static case. This applies, of course, only when the pair of permittivities $\varepsilon$, $\varepsilon_c$ falls in “allowable” regions in Fig. 2. As a special case of interest, we may consider the case in which the core sphere is perfectly conducting, for which $\varepsilon \to i\infty$. In this case, we get $a = \sqrt[3]{(\varepsilon_0 - \varepsilon_c)/(2\varepsilon_c + \varepsilon_0)}\, a_c$, which may be satisfied when $\varepsilon_c$ is chosen in the interval $0 < \varepsilon_c < \varepsilon_0$.

It may be noted here that with some other combinations of parameters it is possible to derive another ratio $\gamma$ for which the determinant $V_n^{TM}$ in (6) becomes zero. As already discussed in our previous work [21]-[22], this ratio would induce an anomalous plasmonic resonance in the core-shell system, causing an opposite effect, namely a highly increased scattering from an electrically small core-shell system.

When the size of the sphere increases, expression (9) is no more effective in determining the exact radius of the outer core to achieve the transparency condition. In this case, however, it is still possible to achieve another condition for which $U_1^{TM} = 0$, which happens at the particular $\gamma$ that makes the determinant in (2) zero. This condition is effective for bigger dimensions of the sphere for which the quasi-static limit does not apply, even though this does not necessarily imply that the scattering from the sphere will become zero, since $U_1^{TM} = 0$ is for the dipolar term $c_1^{TM}$ to be zero, and for too large spheres higher order terms in (4) may become more significant than the first-order term. In this case, however, it may be possible to exploit a similar approach with a multi-layer cover, which should provide more degrees of freedom, and thus a proper design might make simultaneously zero some of the higher-order terms that contribute noticeably to the scattering of the particle. This is one of the subjects in this area we are currently investigating.

It is interesting to note that it may be possible to derive analogous theoretical results for other canonical geometries. In the cylindrical case, for instance, following similar steps it may be shown that the "transparency" condition for the TE (with respect to the cylinder axis, i.e., with the magnetic field parallel with the cylinder axis) polarization becomes:

$$\begin{array}{ll} \gamma = \sqrt[2n]{\dfrac{(\varepsilon_c - \varepsilon_0)(\varepsilon_c + \varepsilon)}{(\varepsilon_c - \varepsilon)(\varepsilon_c + \varepsilon_0)}} & \text{for } n \neq 0 \\ \gamma = \sqrt{\dfrac{\mu_2 - \mu_0}{\mu_2 - \mu_1}} & \text{for } n = 0 \end{array}, \tag{10}$$

where $\gamma$ represents again the ratio of core-shell radii, and the TM formulas may be derived by duality. Other geometrical shapes and reference systems would allow obtaining analogous conditions, which are valid in the quasi-static approximation, and may be similarly extended to the full-wave analysis.

## NUMERICAL RESULTS AND DISCUSSION

An important question is how large the radius $a$ may become in the general full-wave scattering case while one would still be able to achieve a drastic reduction of scattering cross section with a single cover, canceling just the first order term $c_1^{TM}$. For this purpose, Fig. 3 shows the full-wave total scattering cross section of a dielectric particle ($\varepsilon = 4\varepsilon_0$) covered by a plasmonic or metamaterial cover with $\varepsilon_c = -3\varepsilon_0$, in terms of the ratio $a/a_c$, for different sizes of the core-shell particle. For this combination of parameters the value of $T_{TM}$ in (8) for $n=1$ is given by $T_{TM} \simeq 0.61$. As can be seen from this Figure, when the outer radius of the covered particle is extremely small ($a_c = \lambda_0/100$, $\lambda_0 = 2\pi/k_0$ being the free-space wavelength), its scattering cross section goes to near zero very close to that ratio. Increasing the particle size ($a_c = \lambda_0/10$), the cross section can still become very small, making the object nearly "invisible", (since the higher order scattering coefficients are still negligible), even though the quasi-static solution is no more adequate and the transparency condition should be obtained directly by equating Eq. (2) to zero, shifting downwards the required value of $\gamma$ for the transparency in this case. Even with a larger outer radius ($a_c = \lambda_0/5$) we may get a very low scattering, but of course the value of $\gamma$ for the minimum scattering has been shifted downwards again.

In order to give a physical insight into this transparency phenomenon, one may think of it in the quasi-static approximation and with the goal of simply reducing the dipolar term. As is well known, the dipolar scattering is the main response related to the polarization vector $\mathbf{P} = (\varepsilon - \varepsilon_0)\mathbf{E}$ induced locally by the local field as a result of exciting electric field. A regular dielectric sphere would show a scattered dipolar field due to the total induced electric dipole moment, which in the quasi-static limit corresponds to the integral of the vector $\mathbf{P}$ over the volume of the sphere. A cover with $\varepsilon_c < \varepsilon_0$, however, would show a local polarization vector $\mathbf{P} = (\varepsilon_c - \varepsilon_0)\mathbf{E}$ in the cover, being anti-parallel to the local electric field in the cover, which after being integrated over the shell volume may cancel the original dipole moment of the core itself. With the proper choice of radius $a_c$ for the cover this cancellation may be complete, and this is what formula (9) heuristically represents. This is summarized in Fig. 4 (a similar intuitive explanation has been provided in [23] for the static scenario).

We note that this cancellation phenomenon does not rely on a resonant phenomenon, unlike the opposite phenomenon of the scattering resonance by such a core-shell system (for which $V_n^{TM}=0$) [21]-[22]. This is an interesting point, since it ensures: a) that this phenomenon is not strongly sensitive to losses or to the spherical symmetry of the particle, and therefore ohmic losses or imperfections in the particle shape will not precipitously alter this phenomenon, and b) that the phenomenon may not be strongly sensitive to the ratio condition (8) (i.e., it has a larger "ratio bandwidth"), as can also be seen in the examples of Fig. 3.

Using a similar argument, we may apply the transparency phenomenon to a plasmonic particle ($\varepsilon=-3\varepsilon_0$, $a=\lambda_0/10$) by coating it with a dielectric material ($\varepsilon_c=10\varepsilon_0$). For this combination $T_{TM}\simeq 0.825$ and therefore the required radius for the cover is $a_c\simeq 1.21a$ in order to achieve transparency. This is shown in Fig. 5, where the scattering cross section (a) and the contributions from the several scattering coefficients (b) are depicted varying the cover radius. It is interesting to notice how close to the "near-zero-scattering" ratio we find a peak in the scattering cross section (Fig. 5a), which is clearly due to the resonance of the $c_2^{TM}$ scattering coefficient (Fig. 5b). This is expected when $V_2^{TM}=0$ in Eq. (3), which in the quasi-static approximation is expected for $a_c\simeq 1.05a$. It can also be noticed that the transition in which $c_1^{TM}$ approaches zero in Fig. 5b is a relatively smooth one (very different form the scattering peak, as discussed in [21]), implying that the range of the ratio of radii in Fig. 5a for which the scatterer has a very low scattering cross section is relatively broad. As was just pointed out, this is another indication of the "non-resonant" behavior of this phenomenon.

Fig. 6 reports the same plots as in Fig. 5, but for a larger particle ($a=\lambda_0/5$). The results show that even for a particle far from the quasi-static condition, the scattering cross section may be drastically reduced with the same principle. (One may appreciate how the higher order terms in (4) are significant in Fig. 6b.) The effectiveness of the cover is due to the fact that the higher order scattering terms are also in general affected when such a "complementary" cover is employed (Fig. 6b).

It should be underlined that the combinations of parameters chosen in Figs. 4, 5 and 6 also allow the presence of the resonance peaks described in [21] (at which the different $V_n$ may go to zero) for other ratios of radii (they are clearly seen in Figs. 5b and 6b). For larger particles the presence of this peak worsens the transparency performance of these plasmonic and metamaterial covers: the minimum dip we see in the plot of scattering cross section is closely surrounded by the regions where the resonant maxima are present, and as a result a "good" transparency condition may not be effectively achieved for larger particles (at least with a single-layer homogeneous cover with negative-parameter metarials). Therefore, in order to achieve better transparency performance for particles that do not satisfy the quasi-static approximation, the transparency ratio $T_{TM}$ and the high-scattering resonant ratios should be as far apart as possible. There is indeed a range of material parameters in Fig. 2 where this condition is possible: when one of the two materials has a positive permittivity, but lower than the permittivity of the outside region (e.g., free space). In this range of material parameters, it can indeed be shown that no resonance peak may arise in the quasi-static approximation (the determinant in (6) cannot be zero for material parameters in this range [21]).

Fig. 7, as an example, shows the results for a sphere with $\varepsilon = 10\varepsilon_0$ and $a = \lambda_0 / 10$ covered with a material with $\varepsilon_c = 0.5\varepsilon_0$, as a function of $a_c / a$. In this case, as can be seen, the reduction of total scattering cross section can also be achieved, and it is clearly evident from Fig. 7b that no high-scattering resonant peak is around this transparency point.

The total scattering cross section of larger objects may be successfully reduced by further reducing the positive permittivity of the cover. Fig. 8 shows some results for the same inner material as in Fig. 7, but twice its radius (and so eight times its volume). The cover utilized now has $\varepsilon_c = 0.1\varepsilon_0$. We note how the scattering cross section has been drastically reduced at the optimum radius for which $c_1^{TM} = 0$.

The residual scattering cross section at the minimum dip in Fig. 8a is mainly due to the contribution of the magnetic dipole contribution $c_1^{TE}$, since this is not directly affected by the low permittivity of the cover, as can be seen from Fig. 8b. Provided that we may find (or synthesize, as discussed in the introduction) a material whose permeability may be brought down towards zero at the same frequency, however, one might be able to further reduce the scattering from the object using a similar principle. Choosing a cover permeability as $\mu_c = 0.025\mu_0$, for which $U_1^{TE} = 0$ at the same $a_c = 1.12a$ where we get the minimum in Fig. 8a, we are able to make the object even less "visible", as shown in Fig. 9.

Fig. 10 shows the contour plots of the magnitude of the radial component of the near-zone scattered electric field for the case of a dielectric sphere with $\varepsilon = 10\varepsilon_0$, $\mu = \mu_0$ and $a = \lambda_0/5$ in four different cases: no cover (Fig. 10a), a dielectric cover with $\varepsilon_c = 5\varepsilon_0$ and $a_c = 1.12a$ (Fig. 10b), the same cover as in Fig. 8 (Fig. 10c) and the same cover as in Fig. 9 (Fig. 10d). In all these cases the system is assumed to be excited by a plane wave propagating along the $\hat{\mathbf{z}}$ direction with a unit-amplitude electric field polarized along $\hat{\mathbf{x}}$, and the plots are in the *x*-*y* plane, where the electric multipole contribution is maximum. We note that the contour plots are all normalized to the same value, so we can compare the field amplitudes in the different figures. As can be easily seen, in the first case without cover (Fig. 10a) the electric field induces a strong electric dipole moment inside the sphere, which consequently generates a dipolar scattered field in the outside region. This is the usual situation. Adding a cover made of a regular dielectric with arbitrarily selected $\varepsilon_c = 5\varepsilon_0$ the scattered field is further enhanced (Fig. 10b), as expected. The scattered radiation becomes even less symmetric due to the contribution of the quadrupole and higher order multipoles. However, when the cover used in Fig. 8 with $\varepsilon_c = 0.1\varepsilon_0$, $\mu_c = \mu_0$ and $a_c = 1.12a$ is used (Fig. 10c), a scattered field comparable with the previous cases (a and b) is observed inside the object, but the cover almost cancels the scattered radiation outside. The residual scattered radiation (which is very weak compared with the scattered radiation in Fig. 10a or 10b) is clearly due to the quadrupolar and octupolar terms, but it has been drastically reduced in magnitude, consistent with what was shown in Fig. 8a. When the cover of Fig. 9 (with $\varepsilon_c = 0.1\varepsilon_0$, $\mu_c = 0.025\mu_0$ and $a_c = 1.12a$) is employed, the scattered field is even further reduced, even though the effect of this cover is mainly seen in the magnetic field plots (Fig. 11), since it has been designed to reduce the magnetic dipole radiation.

Fig. 11 shows the corresponding plots similar to Fig. 10, but for the magnitude of the radial component of the near-zone scattered magnetic field, in the *y*-*z* plane where the magnetic multipoles contribute maximally to the radial components. Fig. 11a shows the radial components of the magnetic field scattered by a single dielectric sphere, and it is dominated by the magnetic dipole contribution. Increasing the size by adding another dielectric cover, as shown in Fig. 11b, produces an increase in the scattered field, and now the magnetic quadrupole becomes also noticeable. When the cover of Fig. 10c is used here, instead, the magnetic near field is similar to the one obtained without any cover, as expected from the results of Fig. 8. However, adding the low-permeability condition to the cover, as shown in Fig. 11d, the magnetic dipole contribution to the scattered field is essentially canceled. Now the object is almost "invisible" in both the E and H planes, i.e., in the whole space.

Fig. 12 shows the *total* time-averaged Poynting vector distribution in the *x*-*z* plane for the same four cases. When the "transparency" covers are used (Figs. 12c, 12d), one notes how the incoming plane wave seems to "pass through the sphere" without being affected noticeably by the scattered field. In other words, an observer seated outside the covered sphere, even in the near field, would not be able to essentially "sense" the presence of the sphere (since there is almost no scattered field outside), although the fields are noticeably changed inside the sphere and its cover. If a conventional scatterer is placed behind such a "transparent" sphere, an observer would interestingly "notice" the presence of that scatterer without essentially "seeing" the "transparent" sphere between him/her and the scatterer itself.

Fig. 13 reports the similar results as in Fig. 10, but for a smaller sphere ($a = \lambda_0 / 20$). In this case, the scattered fields from the sphere alone and from the sphere covered with a standard dielectric (Figs. 13a and 13b) are dominated by the electric dipole term, and the magnetic dipole contribution is negligible. Covering it with the proper cover (since we are in this quasi-static case we may simply use the approximate expression (9), which yields for the radius $a_c = 1.09a$), the transparency is effectively achieved. Due to its small dimensions, there is no need to employ low-$\mu$ materials, since the H-plane field contribution is already negligible. The related plots therefore are not reported here.

We need to point out the sensitivity/robustness of the results presented here to some realistic parameters. As discussed previously, the transparency phenomenon is not due to a resonance, but just relies on the overall cancellation of the multipolar scattering fields (see Fig. 4). This is why this transparency effect has a relatively broader range for selection of ratio of radii and cover sizes (we are not restricted to one single value for the cover, but there is a relatively broader range of values for which a cover with negative polarizability may "hide" a dielectric sphere). By the same token, this phenomenon may be robust with the variation of the material losses.

Fig. 14, as an example, shows how the results of Fig. 7a and Fig. 8a are modified by the presence of losses in the cover material. As can be seen, the transparency effect is not affected much by the presence of reasonable

ohmic (material) losses. For the dispersion with the Drude or Lorentzian models, in the frequency regions near the plasma frequencies where the constitutive parameters have values close to zero (which we have shown to be more effective in inducing this “transparency” effect) there are no sharp resonances in the medium dispersion, and therefore, following the Kramers-Kronig relations, we expect less ohmic losses than near the resonant frequencies [4].

We speculate that this setup is not too sensitive to possible imperfections in the construction and geometry of the cover. Being a non-resonant phenomenon, we do not expect substantial changes in the results reported here for quasi-spherical or quasi-cylindrical objects with small imperfections on their surface. The fact that a negative polarizability material would effectively cancel one or more multipolar contributions of the scattered field is indeed not a property of the spherical or the cylindrical geometry, and it is predictable that formulas analogous to (9) and (10) may be derived as well for several other geometries and for more complex objects. As a final remark, we should point out that this transparency phenomenon may somehow be related to the issues of non-radiating sources in the inverse scattering and source problems in electromagnetics and acoustics [24]-[25]. We are currently investigating this point.

## SUMMARY

In this contribution we have studied how it is possible to drastically reduce the scattering cross section of spherical and cylindrical objects using lossless plasmonic or metamaterial covers. We have provided physical insights and numerical examples of how a proper design of these metamaterial covers near their plasma resonance may induce a dramatic drop of the scattering cross section (even in the absence of material loss) making the object nearly “transparent” to an external observer. We have also discussed how this effect is insensitive to the possible losses or other imperfections in structures. This phenomenon can provide numerous potential applications in the design of low-observable targets, low-coupling in densely-packed devices, and non-invasive field nano-probes. These issues are currently under investigation.

## ACKNOWLEDGEMENTS

This work is supported in part by the U.S. Defense Advanced Research Projects Agency (DARPA) Grant Number HR0011-04-P-0042. Andrea Alù has been partially supported by the 2004 SUMMA Graduate

Fellowship in Advanced Electromagnetics. We thank Dr. W. Ross Stone for pointing us to the references [24] and [25] and the topic of inverse scattering problems and Dr. Alexey Vinorgradov for bringing to our attention Ref [17].

**FIGURES**

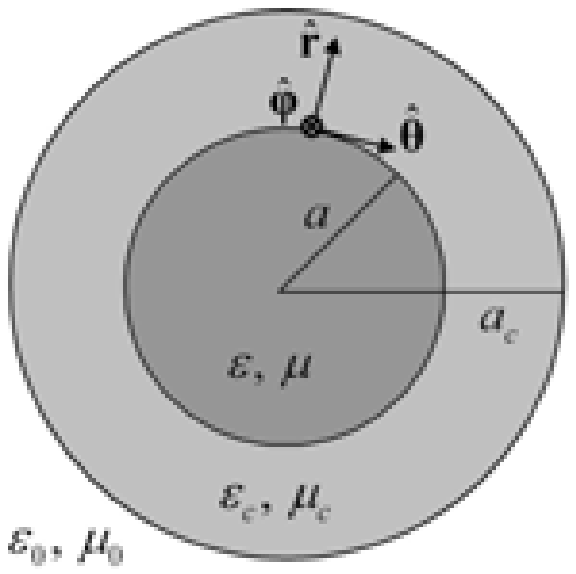

Figure 1 – Cross section of a spherical scatterer composed of two concentric layers of different isotropic materials in a suitable spherical reference system.

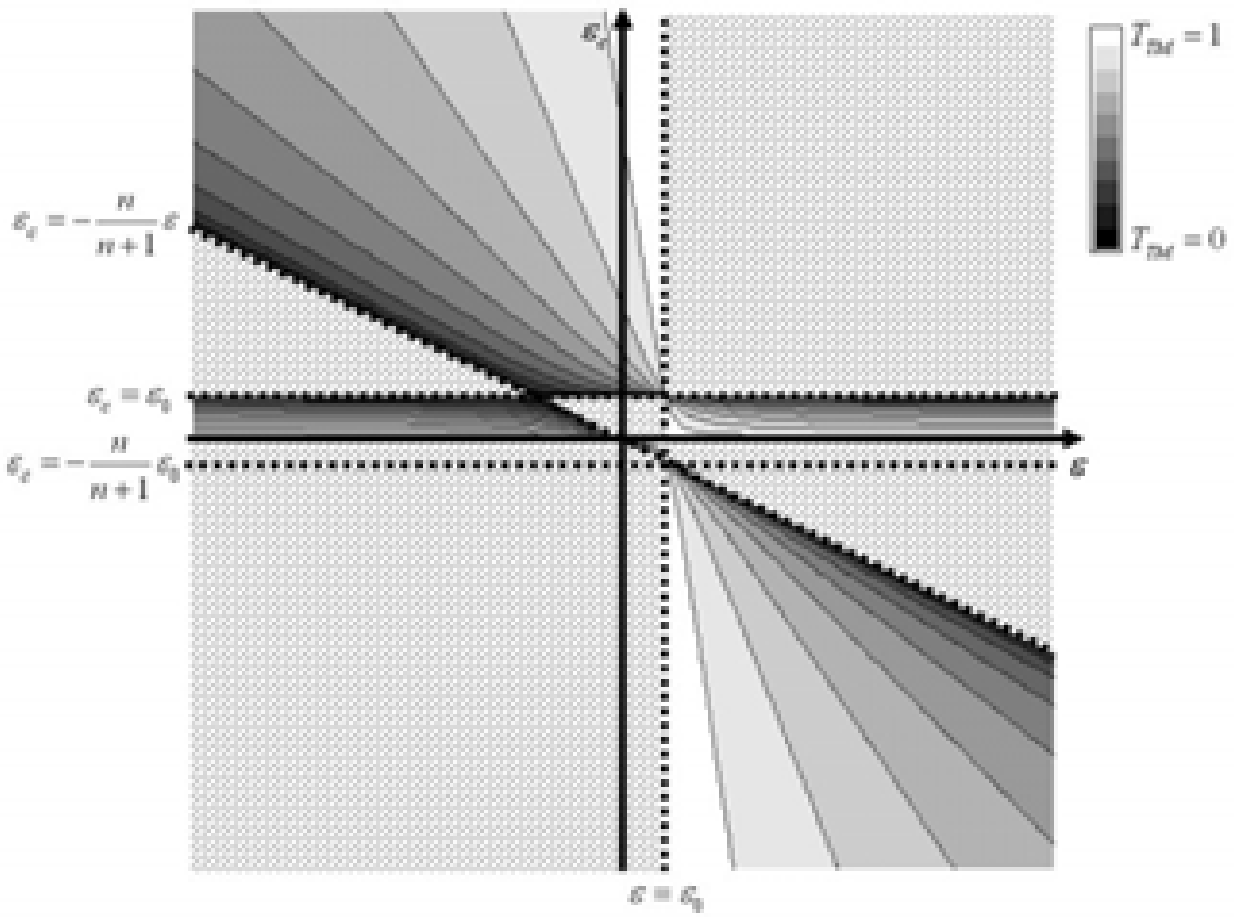

Figure 2 – Ranges of material permittivities for which the condition (8) admits physical solutions and contour plot for the values of the ratio $T$ (TM polarization).

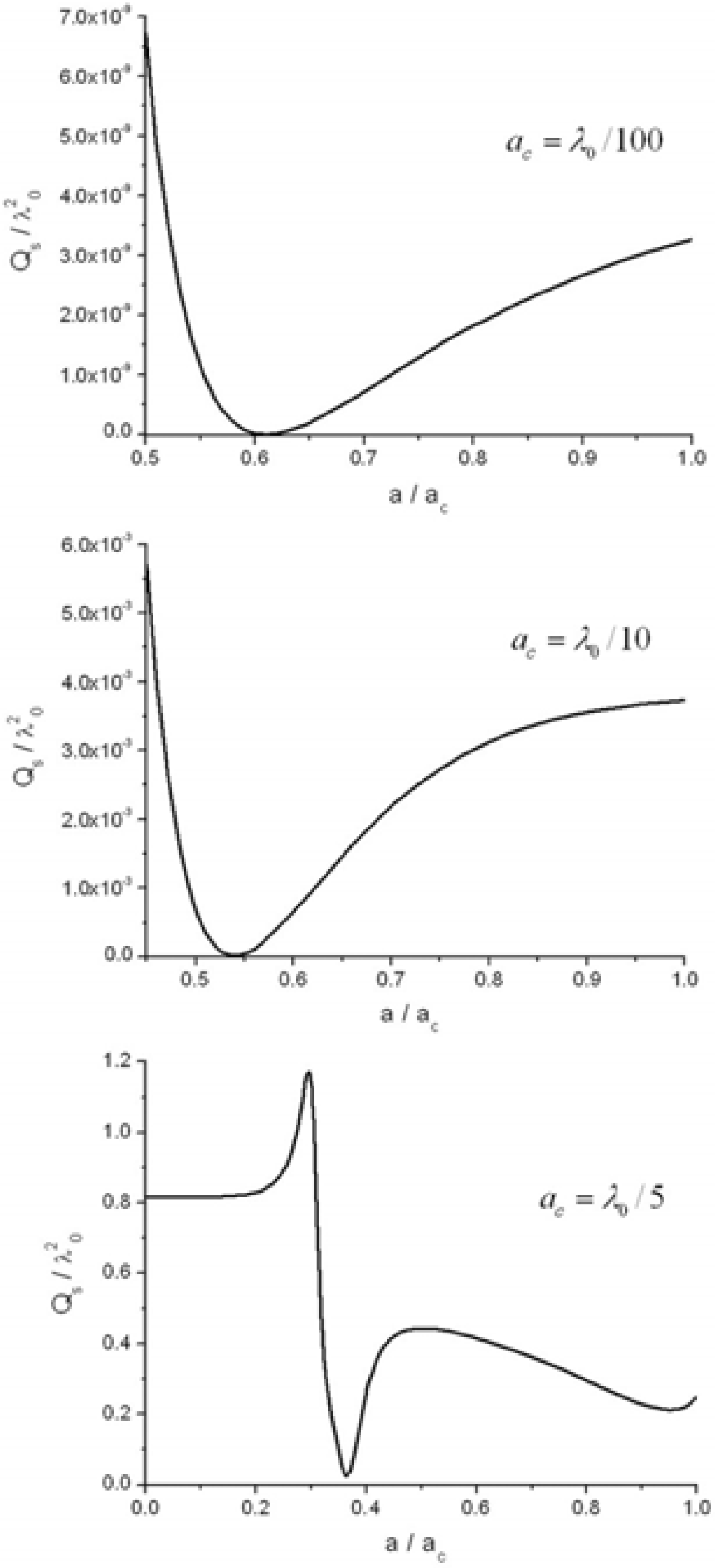

Figure 3 – Normalized scattering cross section for a spherical particle with $\varepsilon = 4\varepsilon_0$, $\mu = \mu_0$, and the cover with $\varepsilon_c = -3\varepsilon_0$ , $\mu_c = \mu_0$, versus the ratio $a / a_c$ for three different sizes of the outer radius of the cover.

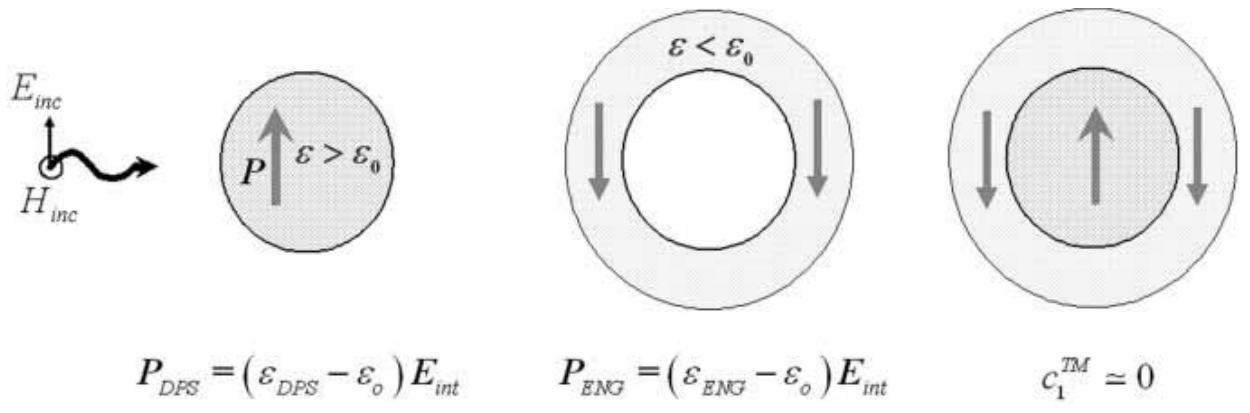

Figure 4 – Heuristic interpretation of the transparency phenomenon: cancellation of the overall dipole

moment through an induced negative polarization vector.

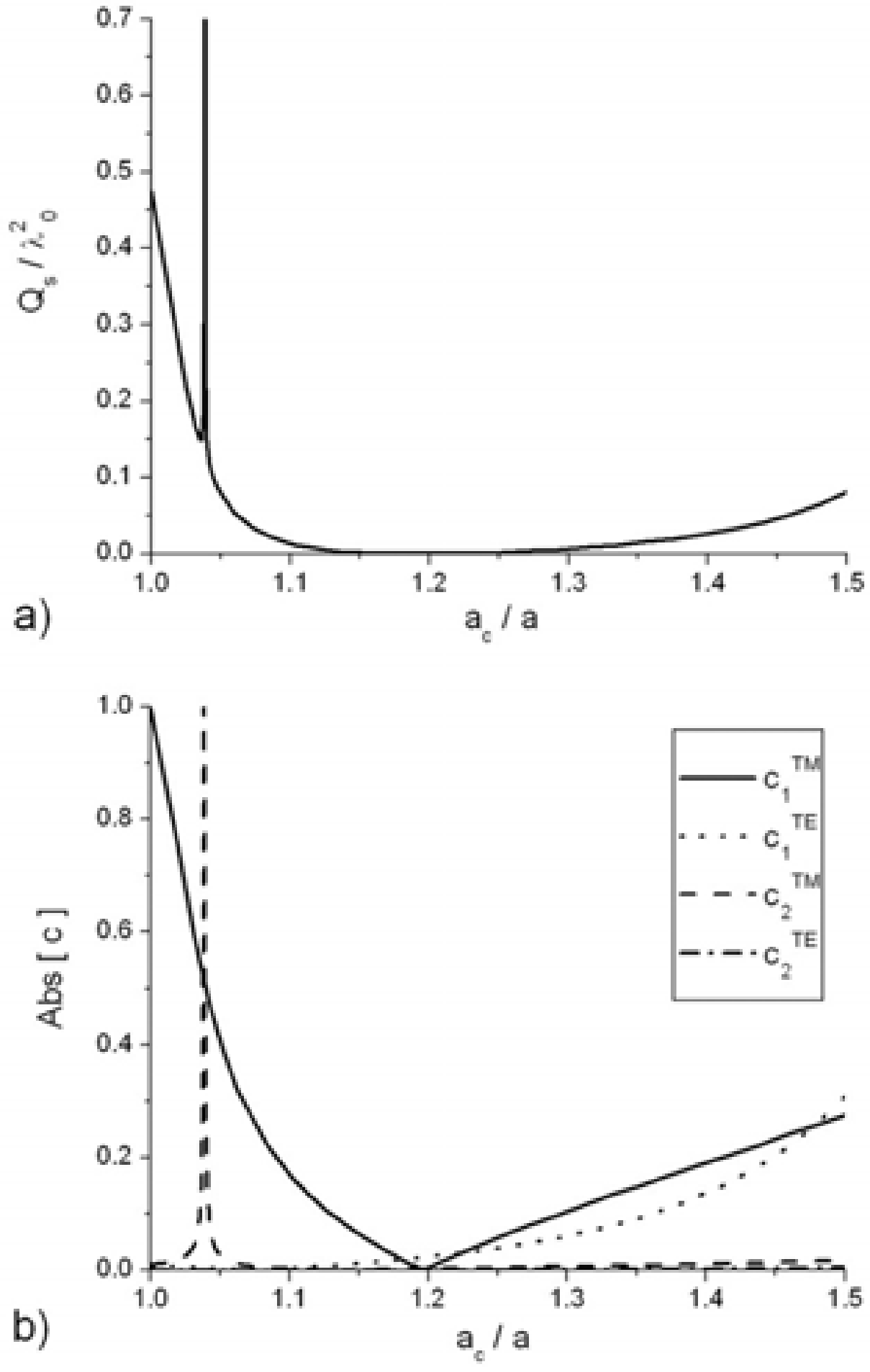

Figure 5 – (a) Normalized total scattering cross section and (b) contributions of the several (non-zero) scattering coefficients, in terms of the cover radius for a fixed inner core ($\varepsilon = -3\varepsilon_0$, $\varepsilon_c = 10\varepsilon_0$, $\mu = \mu_c = \mu_0$, $a = \lambda_0/10$).

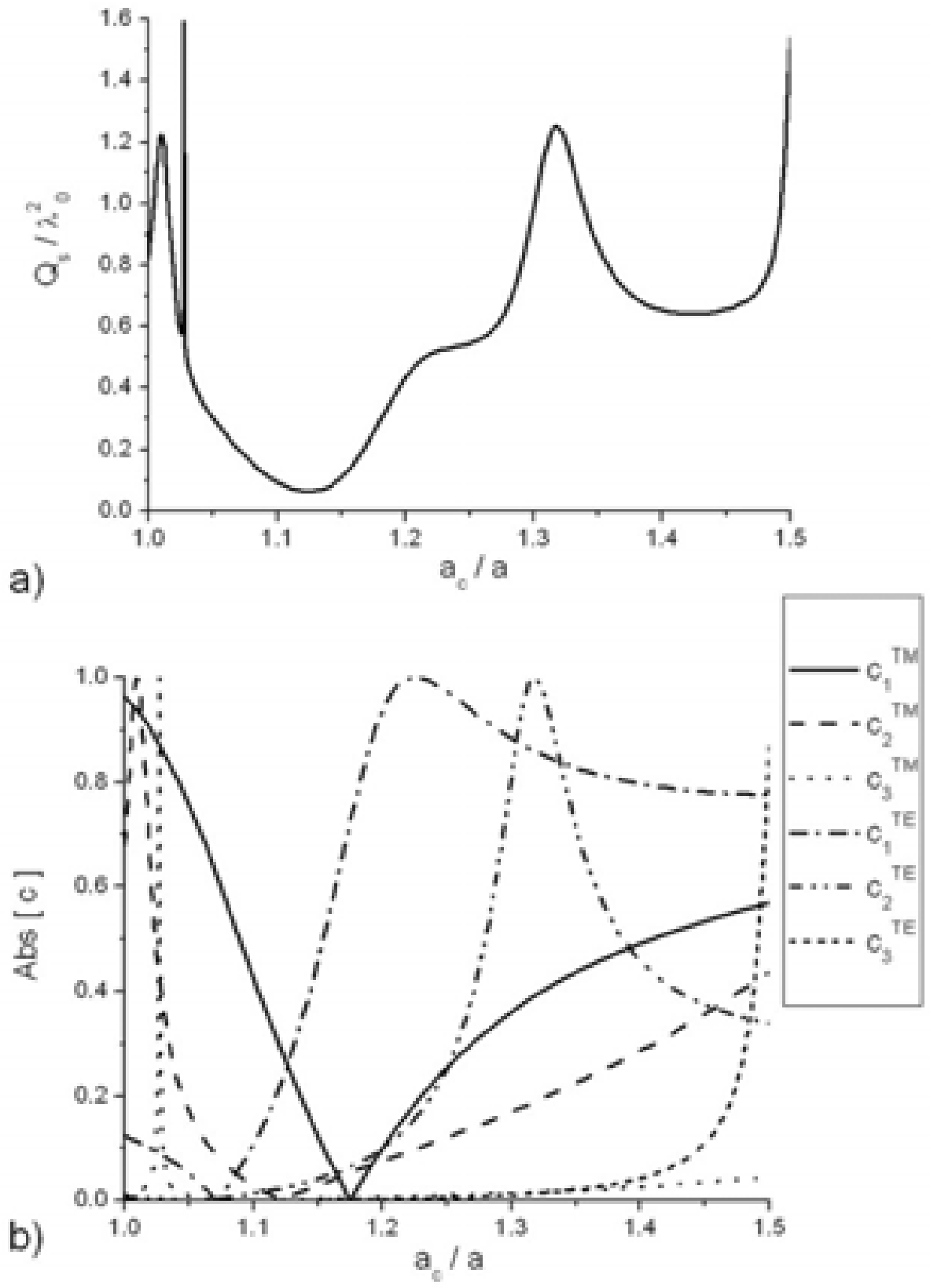

Figure 6 – The same as Figure 5, but $a = \lambda_0 / 5$ and higher order terms become significantly higher. But still, a significant reduction of the scattering cross section may be achieved.

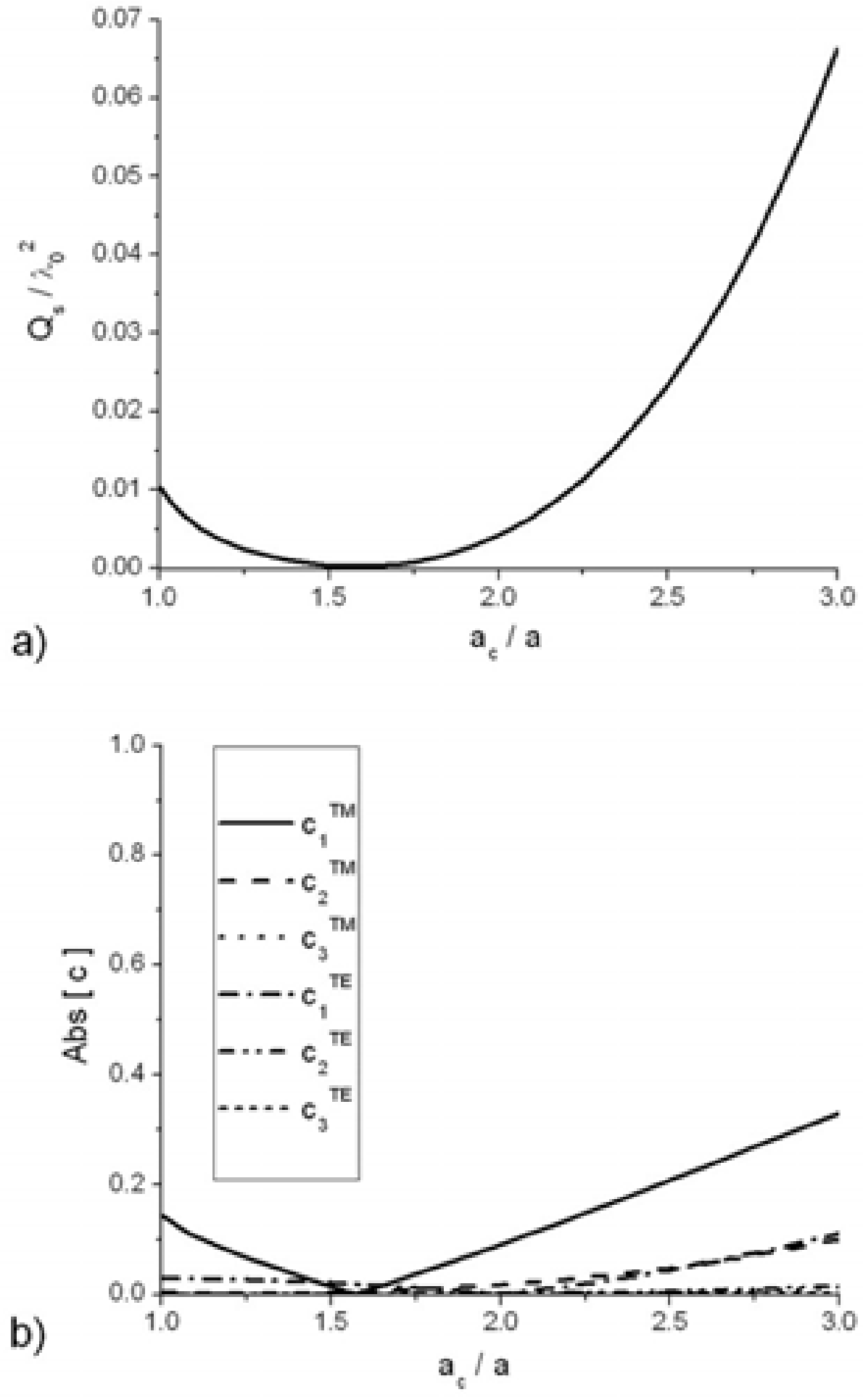

Figure 7 – (a) Normalized total scattering cross section and (b) contributions of the several (non-zero) scattering coefficients, in terms of the cover radius for a fixed inner core ($\varepsilon = 10\varepsilon_0$, $\varepsilon_c = 0.5\varepsilon_0$, $\mu = \mu_c = \mu_0$, $a = \lambda_0 / 10$).

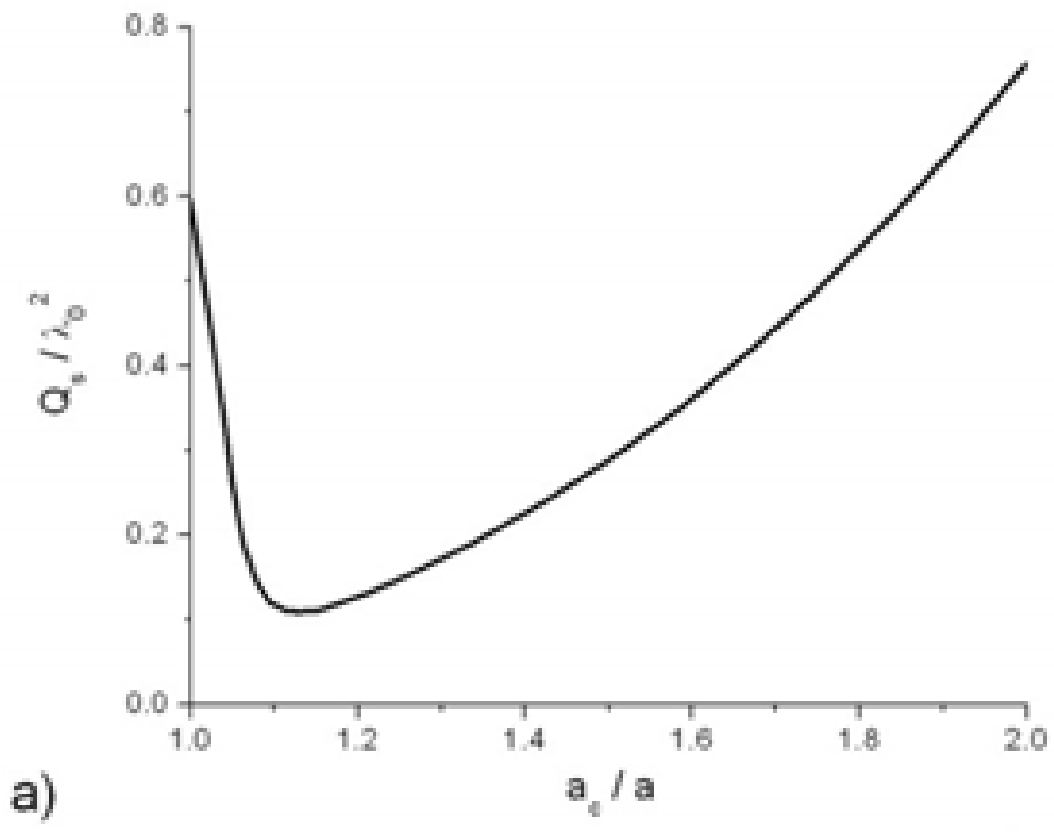

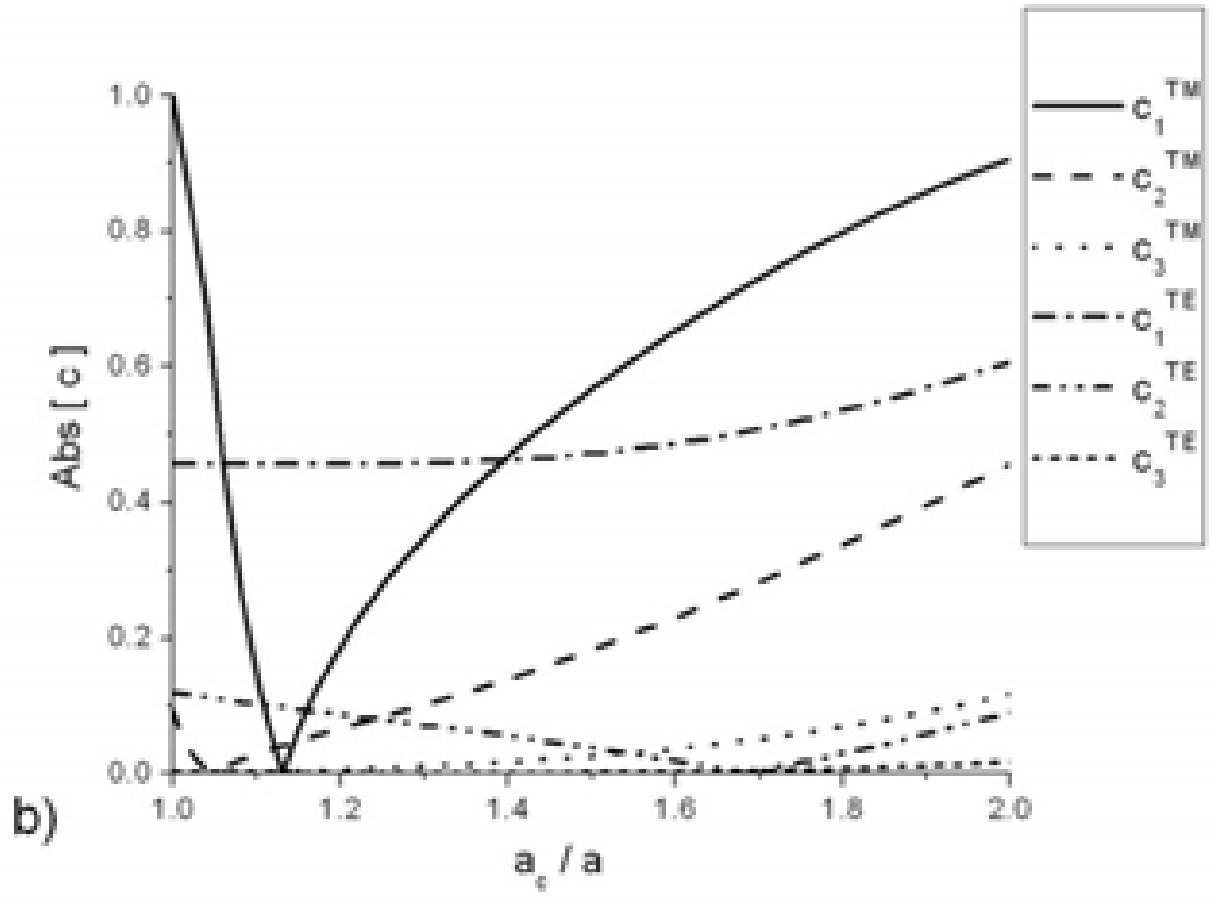

Figure 8 – (a) Normalzied total scattering cross section and (b) contributions of the several (non-zero) scattering coefficients, in terms of the cover radius for a fixed inner core ( $\varepsilon = 10\varepsilon_0$ , $\varepsilon_c = 0.1\varepsilon_0$ , $\mu = \mu_c = \mu_0$ , $a = \lambda_0 / 5$ ).

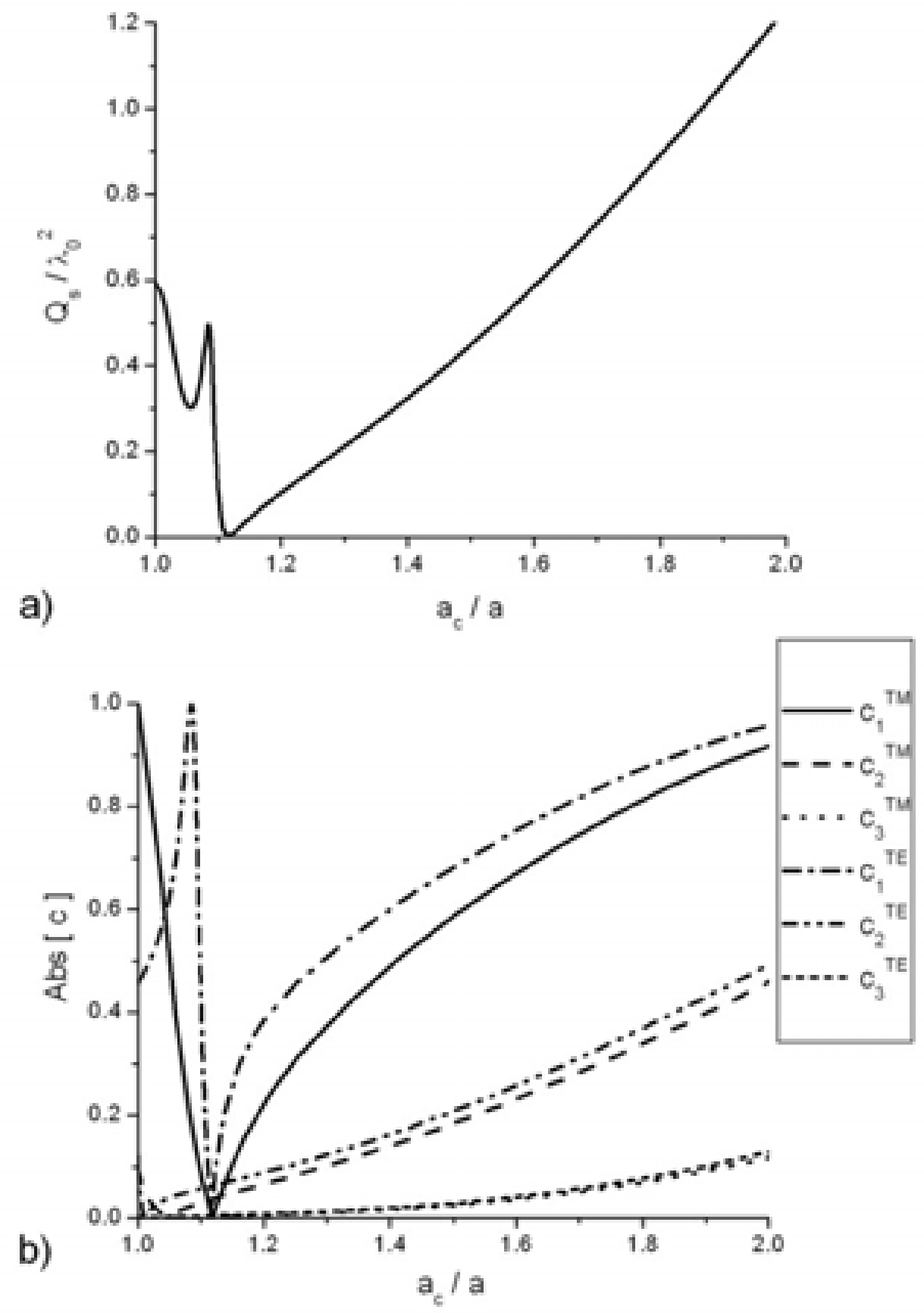

Figure 9 – (a) Normalized total scattering cross section and (b) contributions of the several (non-zero) scattering coefficients, in terms of the cover radius for a fixed inner core ($\varepsilon = 10\varepsilon_0$, $\varepsilon_c = 0.1\varepsilon_0$, $\mu = \mu_0$, $\mu_c = 0.025\mu_0$, $a = \lambda_0 / 5$).

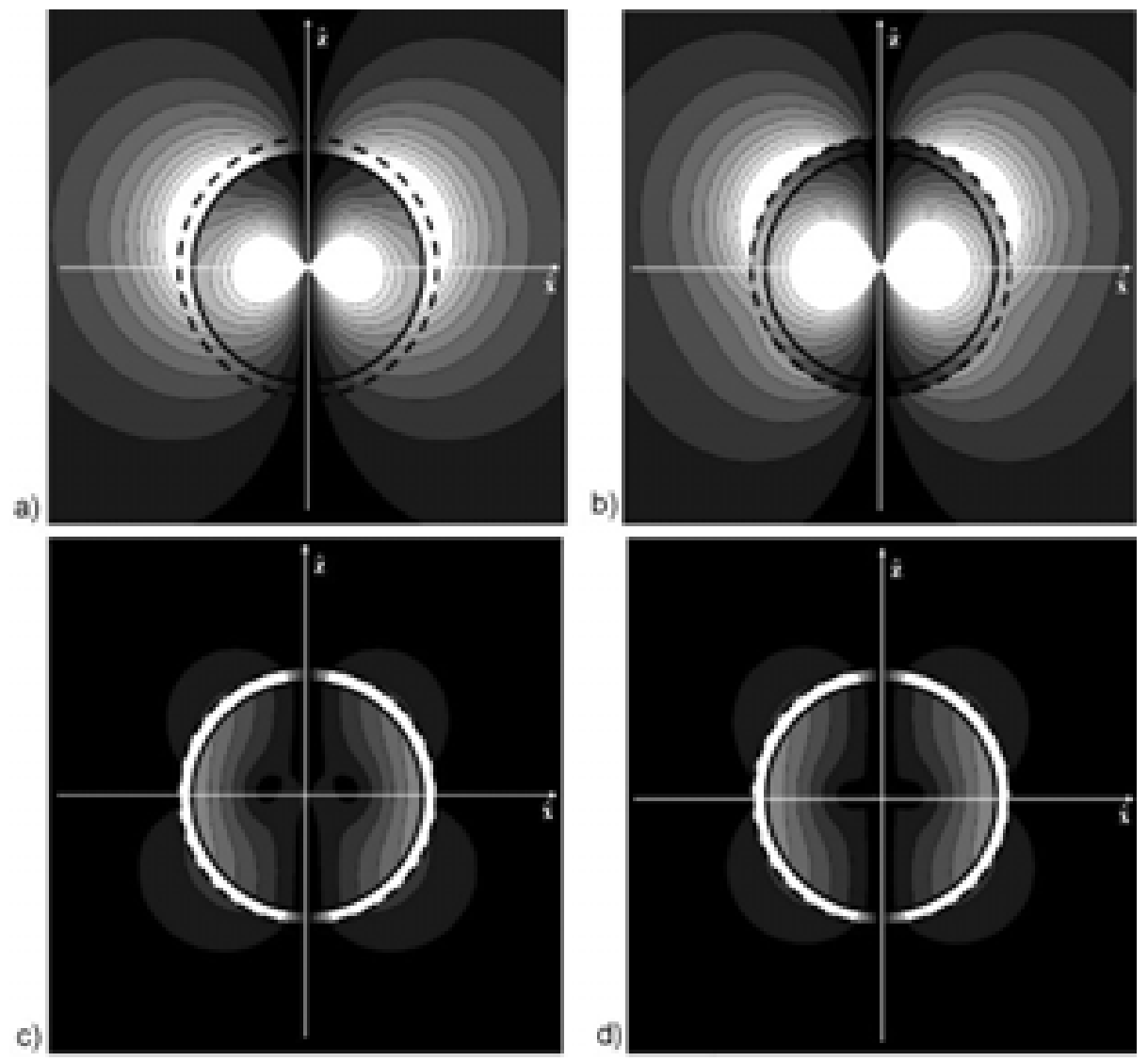

Figure 10 – Contour plots of the distribution of the magnitude of radial component of the *scattered* electric field in the x-z plane, induced by a plane wave traveling along the *z* direction with an electric field polarized along the *x* axis: (a) for a sphere with $\varepsilon = 10\varepsilon_0$, $\mu = \mu_0$, $a = \lambda_0 / 5$; (b) for the same

sphere, but covered with $\varepsilon_c = 5\varepsilon_0$, $\mu_c = \mu_0$, $a_c = 1.12a$; (c) the same as (b), but with $\varepsilon_c = 0.1\varepsilon_0$; (d) the same as (c), but with $\mu_c = 0.025\mu_0$.

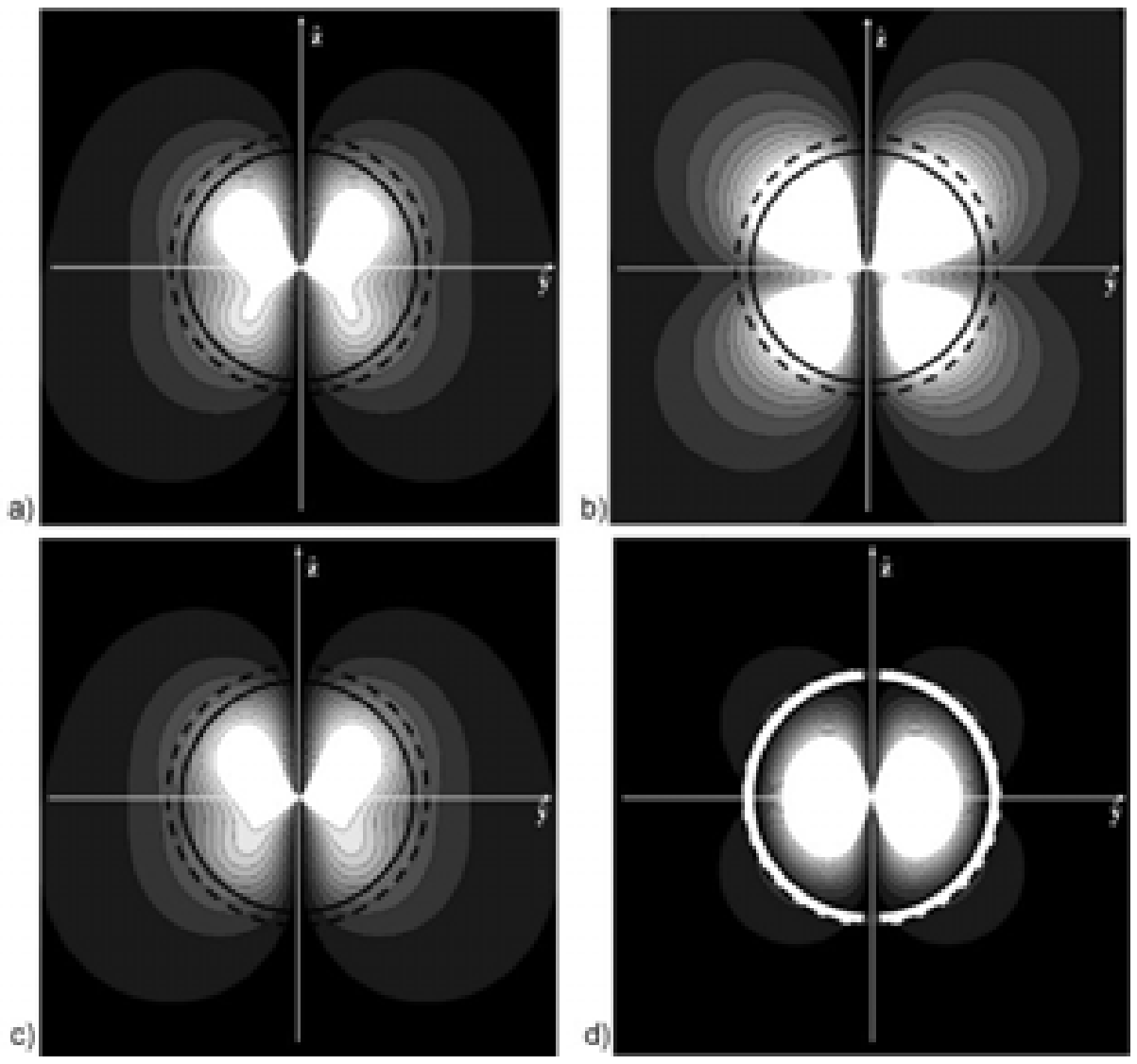

Figure 11 – Contour plots of the distribution of the radial component of the *scattered* magnetic field in the y-z plane, analogous to Fig. 10.

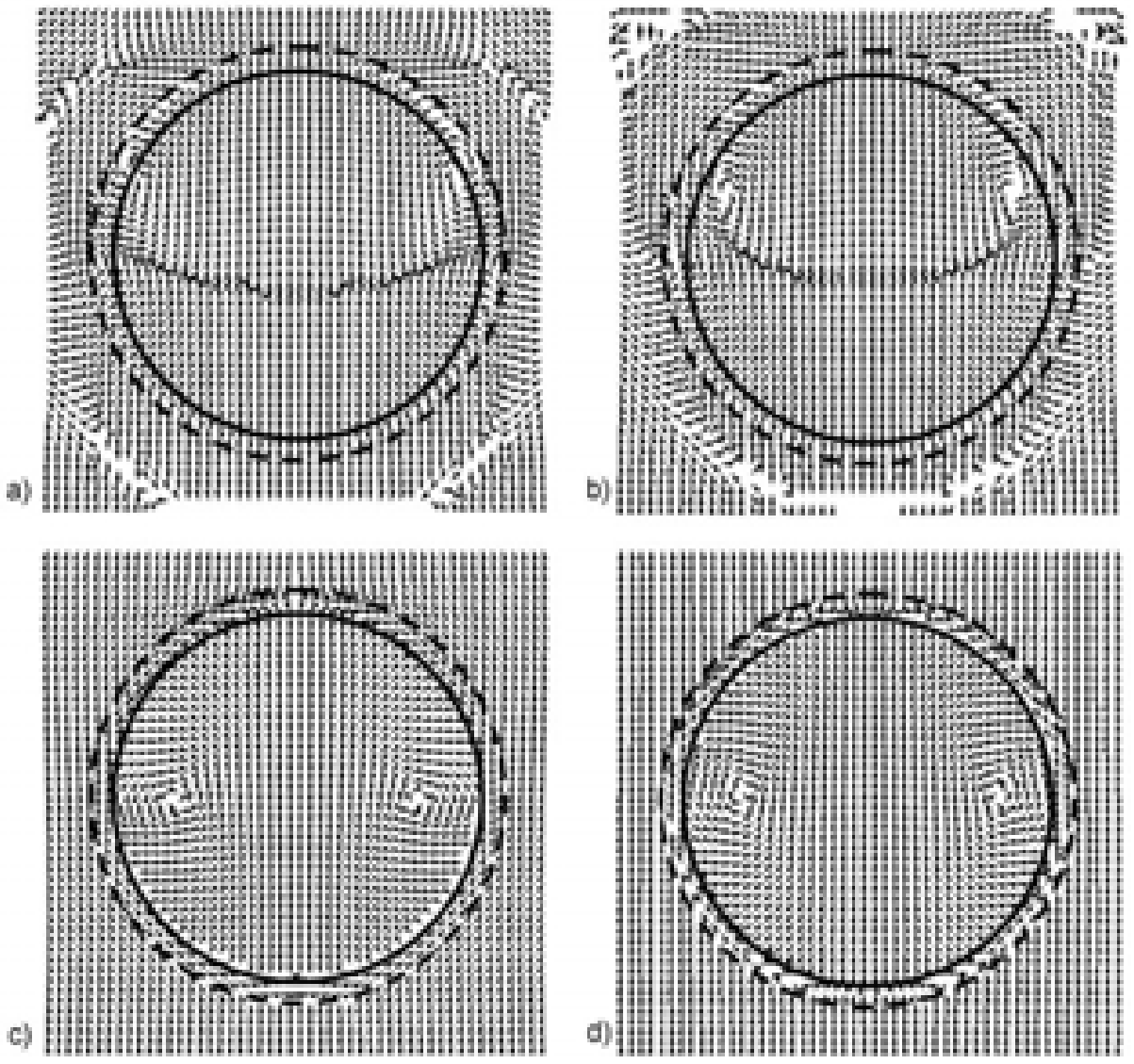

Figure 12 – Total time-averaged Poynting vector distributions in the *x-z* plane for the cases of Figs. 10-11.

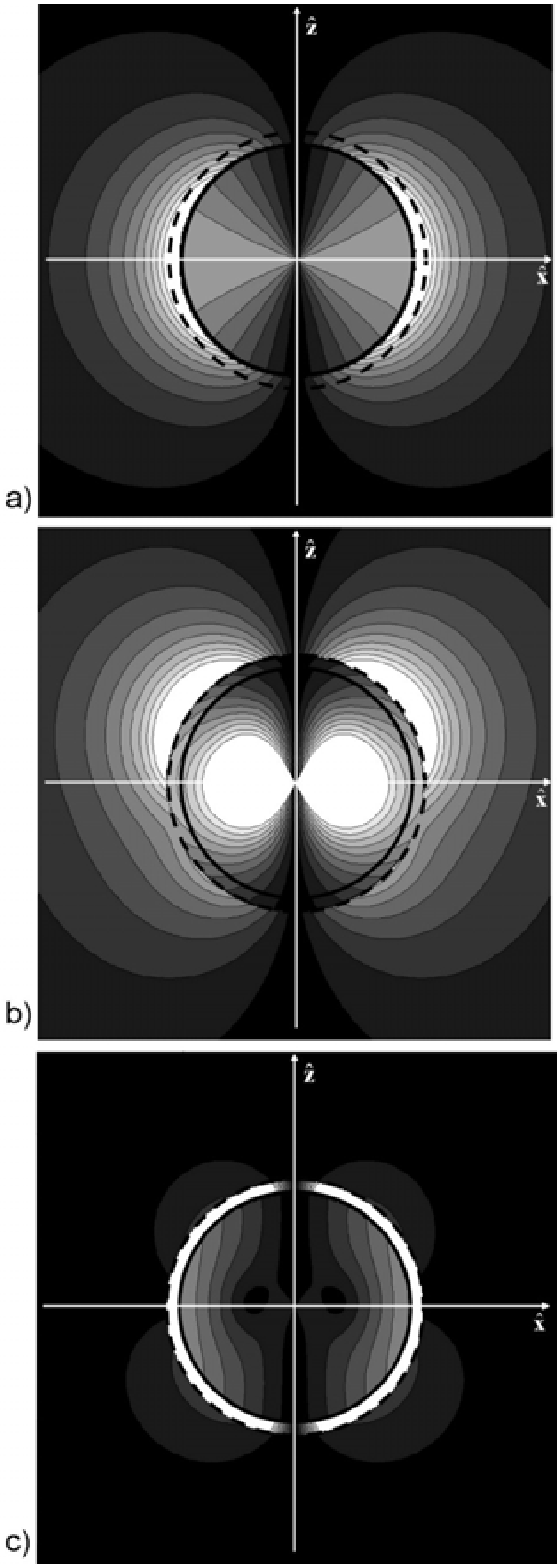

Figure 13 – Contour plots of the distribution of the radial components of the scattered electric field in the x-z plane, induced by a plane wave traveling along the *z* axis with an electric field polarized along the *x* axis: (a) for a sphere with $\varepsilon = 10\varepsilon_0$, $\mu = \mu_0$, $a = \lambda_0 / 20$; (b) for the same sphere, but covered with $\varepsilon_c = 5\varepsilon_0$, $\mu_c = \mu_0$, $a_c = 1.09a$; (c) the same as (b), but with $\varepsilon_c = 0.1\varepsilon_0$.

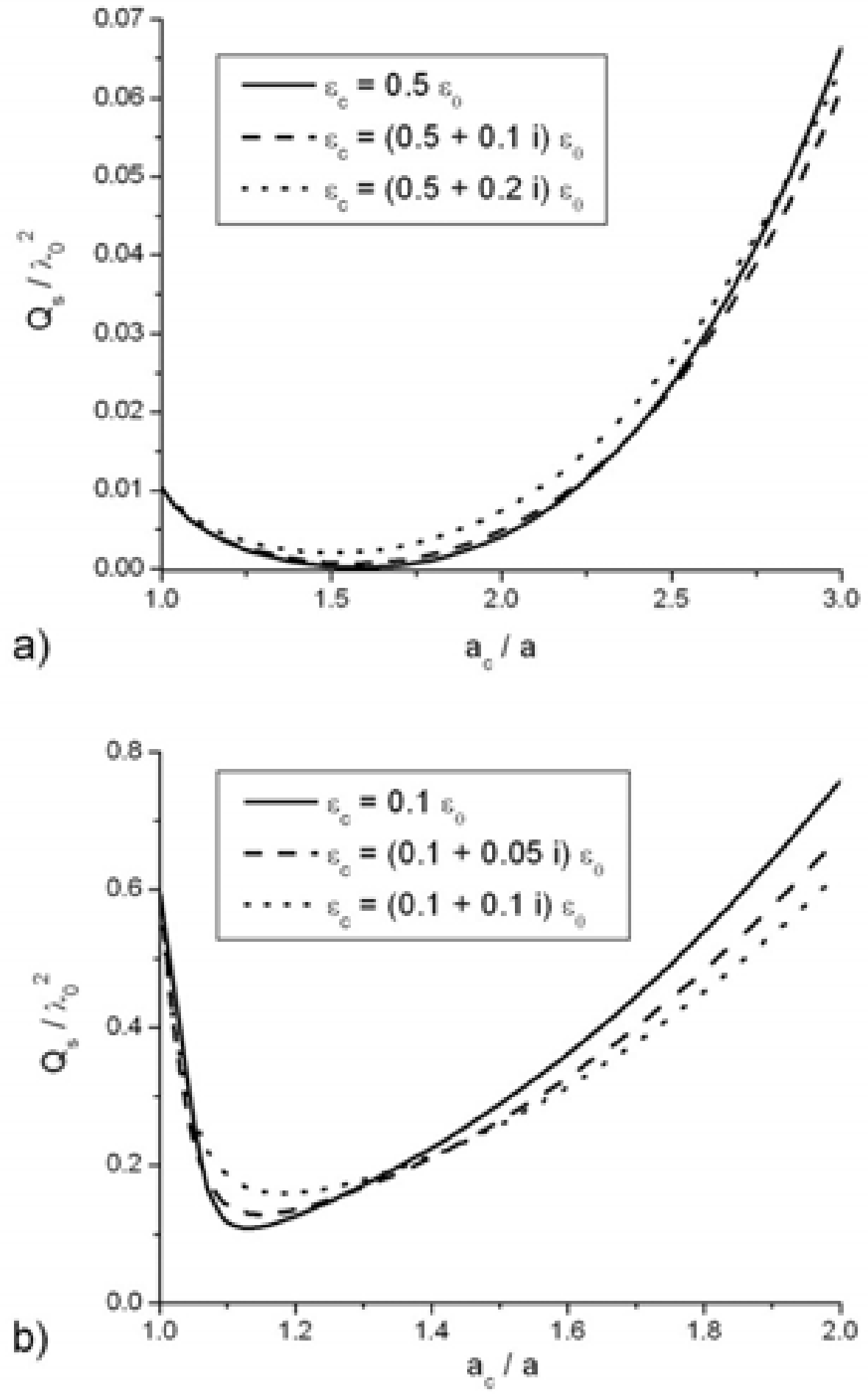

Figure 14 – (a) The same as Fig. 7a; and (b) the same as Fig. 8a, considering ohmic losses in the cover materials.